\documentclass{article}  
\usepackage{natbib}
\usepackage{amssymb}

\begin{document}

\title{A Devastating Example for the Halfer Rule\thanks{This paper appears
    in {\em Philosophical Studies}, Volume 172, Issue 8, pp, 1985-1992,
    August 2015.  The final publication is available at Springer via http://dx.doi.org/10.1007/s11098-014-0384-y}}
\author{Vincent Conitzer\\Duke University}
\date{}

\maketitle
\begin{abstract}
  How should we update {\em de dicto} beliefs in the face of {\em de se}
  evidence?  The Sleeping Beauty problem divides philosophers into two
  camps, {\em halfers} and {\em thirders}.  But there is some disagreement
  among halfers about how their position should generalize to other
  examples.  A full generalization is not always given; one notable
  exception is the {\em Halfer Rule}, under which the agent updates her
  uncentered beliefs based on only the uncentered part of her evidence.  In
  this brief article, I provide a simple example for which the Halfer Rule
  prescribes credences that, I argue, cannot be reasonably held by anyone.
  In particular, these credences constitute an egregious violation of the
  Reflection Principle.  I then discuss the consequences for halfing in
  general.
\end{abstract}

\section{Introduction}

It is far from a settled matter how {\em de dicto} beliefs should be
updated when we obtain {\em de se} information.  The Sleeping Beauty
problem is particularly effective at bringing out conflicting intuitions.
In it, Beauty participates in an experiment.  She will go to sleep on
Sunday.  The experimenters will then toss a fair coin.  If it comes up
Heads, Beauty will be awoken briefly on Monday, and then put back to sleep.
If it comes up Tails, she will be awoken briefly on Monday, put back to
sleep, again awoken briefly on Tuesday, and again put back to sleep.
Essential to the problem is that Beauty will be unable to distinguish any
of these three possible awakenings (Monday in a Heads world, Monday in a
Tails world, and Tuesday in a Tails world).  In particular, when being put
back to sleep after a Monday awakening, Beauty will be administered a drug
that prevents her from remembering this awakening, but otherwise leaves her
brain unaffected.  The experiment will end on Wednesday, when Beauty will
be finally awoken in a noticeably different room, so that there is no risk
of her mistaking this event for one of the brief awakenings.  Beauty is at
all times fully informed of these rules of the experiment.

Now, when Beauty finds herself in one of the brief awakening events, what
should be her credence (subjective probability) that the coin has come up
Heads?  Thirders believe that the correct answer is $1/3$, which would be
the long-run fraction of Heads awakenings if the experiment were to be
repeated many times.  Halfers, on the other hand, believe that Beauty's
credence should be unchanged from Sunday, when it should clearly be
$1/2$. One benefit of being a halfer is that being a thirder (or supporting
any fraction other than $1/2$) seems to violate the {\em Reflection
  Principle}~\citep{Fraassen84:Belief,Fraassen95:Belief}: if on Sunday you
are certain that tomorrow, on Monday, you will have credence (say) $1/3$ in
some event, then you should have credence $1/3$ in that event now
already. But (applying what is known as the {\em Principal Principle})
clearly on Sunday the credence in Heads should be $1/2$, because the coin
is fair.  \cite{Elga00:Self} already notes the conflict between thirding
and the Reflection Principle, attributing this observation to Ned Hall, and
considers the Sleeping Beauty problem a counterexample to the Reflection
Principle.

Even if we were certain of the correct answer to the Sleeping Beauty
problem -- presumably, $1/3$ or $1/2$ -- this would fall short of knowing
how {\em de dicto} beliefs should be formed in the face of {\em de se}
evidence in general.  All it would do is place a constraint on how they
should be formed.  Indeed, halfers disagree on how the $1/2$ answer should
generalize to other examples.  But one natural generalization that has been
discussed in several articles~\citep{Halpern06:Sleeping,Meacham08:Sleeping,
  Briggs10:Putting} is the following, called the ``Halfer Rule'' by Briggs.\\

\noindent {\bf The Halfer Rule.}  Determine which possible (uncentered)
worlds are ruled out by the centered evidence; set their probabilities to
zero.  For those that are not ruled out, renormalize the probabilities, so
that they again sum to one while keeping the ratios the same.\footnote{For
  my purposes, it is not necesssary to specify how credences in centered
  worlds are determined, i.e., how the total credence in a possible world
  is divided across its centers.  This is because I will only consider
  credences in uncentered events in what follows.
  \cite{Titelbaum12:Embarrassment} gives an example where halfers obtain an
  implausible credence in a centered event, if a certain condition on how
  the halfer distributes credence across centers holds.}\\

If Beauty adopts the Halfer Rule, she indeed places credence $1/2$ in Heads
after being awoken, because no possible worlds are ruled out.  Again, not
all halfers agree with the Halfer Rule in general.  For example, the Halfer
Rule prescribes that, if Beauty is always told at some point during her
Monday awakening that it is Monday, her credence in Heads at that point
should still be $1/2$, because still no possible world is ruled out.
But~\cite{Lewis01:Sleeping} advocates a version of halfing that results in
a credence of $2/3$ in Heads after being told it is Monday.  This is a
violation of the Reflection Principle -- Beauty knows that she will change
her credence to $2/3$ on Monday, regardless of how the coin came up, and
yet sticks with $1/2$ on Sunday -- and arguably one that is more serious
than the thirder's alleged violation of it, because in this case Beauty
knows where in time she is when her credence is
$2/3$. Indeed,~\cite{Draper08:Diachronic} have pointed out that this
credence of $2/3$ would make Beauty susceptible to a very simple diachronic
Dutch book, where she is sold one bet on Sunday when her credence is $1/2$
and another on Monday when her credence is $2/3$, resulting in a sure loss
overall.\footnote{One may wonder whether, similarly, we could set up a
  Dutch book against the thirder based on her alleged violation of the
  Reflection Principle.  But this would involve her being offered bets on
  Monday awakenings, without being told that it is Monday, but not on
  Tuesday awakenings, and it has been argued that this does not constitute
  a fair Dutch book because the bookie is exploiting information that
  Beauty does not have~\citep{Hitchcock04:Beauty}.  (Also, from being
  offered the bet Beauty might {\em infer} that it is Monday and thereby
  change her credences and decline the bet.)}

More recently,~\cite{Pittard15:When} has also argued against the Halfer
Rule.  As he points out, his own interpretation of halfing can lead to a
disagreement paradox where two participants in an experiment obtain
different credences in spite of having the same information.  (The Halfer
Rule does not lead to this disagreement paradox in his example.) It should
be noted that it would be trivial to turn these disagreeing participants
into a money pump by arbitrage of their different
credences.\footnote{Pittard nevertheless defends these credences, arguing
  that it may be reasonable to consider this a {\em robustly perspectival}
  context, one in which two disputants {\em should} end up having different
  beliefs in spite of them having the same evidence, being able to
  communicate without restriction, etc.  This may be reminiscent of the
  {\em perspectival realism} described by~\cite{Hare10:Realism} (see
  also~\cite{Hare07:Self,Hare09:On}).  \cite{Hare09:On} goes into some
  detail discussing what conclusion two interlocutors, each of whom takes
  herself to be ``the one with present experiences,'' should reach.  If
  indeed they should not be able to reach complete agreement, as seems
  likely, then this would appear to be a robustly perspectival context.
  However, in this case it does not seem possible to turn the situation
  into a money pump, because it does not seem possible to settle any bets
  made in a satisfactory way; we cannot adjudicate from a neutral
  perspective.  Indeed, Hare concludes that the interlocutors should agree
  that the other is correct {\em from the other's point of view}.  In
  contrast, bets made by the participants in Pittard's experiment could
  easily be settled from a neutral perspective.}

In summary, the Halfer Rule is not universally agreed to constitute the
correct generalization of halfing.  On the other hand, it is a very natural
generalization, it has attracted significant support, and it avoids
problems that other interpretations of halfing encounter.  However, I will
now proceed to show that it is fatally flawed.

\section{A Variant with Two Coins}

The Sleeping Beauty variant that I need is very simple.  Beauty will be put
to sleep on Sunday, and be awoken once on Monday and once on Tuesday.  As
always, she will be unable to remember her Monday awakening on Tuesday.
Two fair coins, called ``one'' and ``two,'' will be tossed on Sunday.  When
she wakes up on Monday, Beauty will be shown the outcome of coin toss one.
When she wakes up on Tuesday, she will be shown the outcome of coin toss
two.  Beauty cannot distinguish the two coins, so seeing the outcome of the
coin toss still does not tell her which day it is.  She only learns that
the coin corresponding to {\em today} came up (say) Heads.
Figure~\ref{fi:twocoins} illustrates the example.

\begin{figure}
\begin{center}
\begin{tabular}{|r|c|c|c|c|}\hline
& HH ($1/4$) & HT ($1/4$) & TH (1/4) & TT (1/4) \\ \hline
Monday & see Heads & see Heads & see Tails & see Tails \\ \hline
Tuesday & see Heads & see Tails & see Heads & see Tails \\ \hline
\end{tabular}
\end{center}
\caption{A two-coins variant of the Sleeping Beauty problem with four
  possible worlds, each with probability $1/4$. Note that Beauty is always
  awoken on both days in this variant, but her information upon awakening
  is not always the same.}
\label{fi:twocoins}
\end{figure}

Now consider the following question.  When Beauty is awoken and observes a
(say) Heads outcome, what should be her credence that the coin tosses came
up {\em the same}?  That is, what should be her credence in the event
``(both coins came up Heads) or (both coins came up Tails)''?  It seems
exceedingly obvious that the answer should be $1/2$.  Clearly this was the
correct credence on Sunday before learning anything (by the Principal
Principle), and intuitively, the outcome of the coin toss today -- whatever
it is -- tells Beauty absolutely nothing about whether the coins came up
the same.  This requires that the coins are fair; if each coin had, say, a
$2/3$ chance of coming up Heads, then learning that today's coin has come
up Tails would give Beauty evidence that the coins are less likely to have
come up the same.  But we explicitly assume that the coins are fair.

I will argue in more detail that $1/2$ is the {\em correct} answer shortly.
But, for the reader who is already convinced of that, let me get to the
point and show which credences result from applying the Halfer Rule.  The
possible worlds that are consistent with a Heads observation are HT (coin
one came up Heads and coin two came up Tails), TH, and HH.  Because each of
these three worlds has the same probability {\em ex ante}, applying the
Halfer Rule results in placing credence $1/3$ in each of these worlds.  But
this implies placing only $1/3$ credence in the event that both coins came
up the same, because of the three remaining worlds only HH has them coming
up the same.  By symmetry between Heads and Tails, the Halfer rule also
prescribes $1/3$ credence in the event that both coins came up the same if
Tails is observed.\footnote{Incidentally, applying the Thirder Rule does
  give the right answer: of all Heads awakenings, two are in the HH world,
  in which the coins come up the same, and the remaining two are in the HT
  and TH worlds, in which the coins do not come up the same.  So if we use
  the Thirder Rule, the resulting credence in the event that both coins
  came up the same is $2/4 = 1/2$.  (I apologize for any confusion caused
  by the unfortunate coincidence that the Halfer Rule prescribes $1/3$ in
  this context, and the Thirder Rule $1/2$.)}

\section{The Halfer Rule and the Reflection Principle}

What is so wrong about the Halfer Rule suggesting that the correct credence
is $1/3$ in the above example?  Well, it is now the Halfer Rule that runs
afoul of the Reflection Principle: if Beauty is certain that her credence
on Monday (or, for that matter, Tuesday) will be $1/3$, then why is it not
$1/3$ already on Sunday?  In fact, it seems to me that this violation of
the Reflection Principle is more serious than the thirder's alleged
violation of it in the original Sleeping Beauty problem, for the following
reason.  In the original problem, it would be unreasonable to say that the
fact that the thirder will end up having a credence of $1/3$ {\em on
  Tuesday} implies that she should already have a credence of $1/3$ on
Sunday.  After all, she does not always wake up on Tuesday, and if she were
capable of, in her sleep, recognizing that she has not been awoken, she
would assign credence $1$ in Heads then.  That is why the purported
violation focuses on the Monday credence in Heads, not the Tuesday one.
But it seems illegitimate to consider Monday separately from Tuesday,
because Beauty cannot distinguish them.  Thus, it seems debatable whether
the thirder really violates the Reflection Principle -- more precisely,
whether she violates any version of this principle by which we would care
to abide.  By contrast, in the two-coins example considered here, it does
not seem that the argument that Monday and Tuesday should be considered
together can be of much help to the supporter of the Halfer Rule, because
Beauty is always awoken and, according to the Halfer Rule, always ends up
with a credence of $1/3$.  But I leave formalizing the sense in which the
violation is more serious for another day.

To make matters yet worse for the Halfer Rule, consider the following twist
to the two-coins example.  On both Monday and Tuesday, after Beauty has
observed the coin toss outcome and been awake for a little while longer,
the experimenter tells her what day it is.  Say she observed Heads and was
then told (a bit later) that it is Monday.  Now only two worlds survive
elimination: HH and HT.  The Halfer Rule will assign each of them credence
$1/2$, resulting in a credence of $1/2$ that both coins came up the
same.\footnote{The Thirder Rule still gives $1/2$ as well, because there
  are only two possible centered worlds remaining, namely Monday in HH and
  Monday in HT.}  But this is yet another violation of the Reflection
Principle: after seeing the outcome of the coin toss but before learning
what day it is, Beauty, if she follows the Halfer Rule, places credence
$1/3$ in the event that the coins came up the same, but she also knows that
once she is told what day it is, in either case, she will shift her
credence to $1/2$.  This is perhaps the most egregious violation of the
Reflection Principle that we have encountered, because in this case she is
not put to sleep and does not have memories erased as she transitions from
one credence to another.\footnote{On the face of it, the same happens in
  the Shangri La example given by~\cite{Arntzenius03:Some}. (I thank an
  anonymous reviewer for Philosophical Studies for calling my attention to
  this.)  In this example, someone experiences A or B according to the
  outcome of a coin toss.  He knows, though, that at a certain point in
  time after the experience, any memories of B will be replaced by false
  memories of A, while any memories of A will be left intact, so that he
  will not be able to tell the two cases apart.  Then, while experiencing
  A, he has credence $1$ in Heads, in spite of knowing full well that he
  will later have credence $1/2$ in Heads, without his memory being
  compromised in this particular case.  Of course, this is entirely due to
  the fact that in a parallel case, his memory would be compromised to be
  indistinguishable from what he currently knows will be his (true) memory
  of A.  Thereby, he will lose a piece of information that he currently
  has.  But nothing similar happens in the enriched two-coins example.  At
  the point in time when Beauty is told what day it is, her memory is {\em
    never} compromised, and she {\em never} loses information.}  Again, I
leave formalizing the sense in which the violation is more serious than the
other violations for another day.

\section{What Options Remain for the Halfer?}

If the Halfer Rule is untenable, then is there another full generalization
of halfing that is more defensible?  I have already mentioned a few
interpretations of halfing that do not always agree with the Halfer Rule
and get into their own brands of trouble as a result.  In this final
section, I hope to assess a bit more systematically how halfing may be
generalized in a trouble-free way.

One helpful example to consider is a variant of the two-coins example
introduced earlier.  The only modification that is needed to obtain this
variant is the following.  To cut down on the cost of the various drugs
involved in the awakenings, the experimenter has decided to only awaken
Beauty when the coin corresponding to the current day has come up Heads.
On Tails days, the experimenter just lets her sleep.  Beauty is of course
informed of this modification at the outset.  As a result, on a Heads
awakening it is no longer necessary to show her that the coin has come up
Heads, because this is already implied by the fact that she was awoken at
all.  On the other hand, nothing is lost by showing her the Heads outcome
anyway. Figure~\ref{fi:twocoinsmodified} illustrates the modified example.

\begin{figure}
\begin{center}
\begin{tabular}{|r|c|c|c|c|}\hline
& HH ($1/4$) & HT ($1/4$) & TH (1/4) & TT (1/4) \\ \hline
Monday & see Heads & see Heads & asleep & asleep \\ \hline
Tuesday & see Heads & asleep & see Heads & asleep \\ \hline
\end{tabular}
\end{center}
\caption{A cost-cutting variant of the two-coins example in
  Figure~\ref{fi:twocoins}, the only modification being that Beauty is no
  longer awoken on Tails.}
\label{fi:twocoinsmodified}
\end{figure}

Now what should Beauty believe upon awakening (with Heads)?  It appears to
me that in this variant, any reasonable generalization of halfing must
place credence $1/3$ in each of the worlds HH, HT, and TH.  Specifically,
TT is ruled out by the evidence, HT and TH should have the same credence by
symmetry, and it appears that the only motivation one could have for giving
HH a higher credence is that this world has more centers -- but that is
thirder reasoning!  If these $1/3$ credences are right, it leads to the
following question.  How could the fact that we no longer awaken Beauty on
Tails days affect her correct credence on Heads days?  If one answers that,
well, in fact, it should not affect it, then all is lost for the halfer.
It implies that the halfer is stuck with the Halfer Rule's prescribed
credences for the original two-coins example, which are untenable.

So, the halfer must adopt a position that allows for the prescribed
credence to change when we change whether Beauty is awoken under other
conditions -- conditions that she herself would be able to distinguish from
the current ones.  This may seem unappealing; in particular, the thirder
needs to make no such move.  Still, reasonable generalizations of halfing
may fit the bill.  For example, consider the following approach, based on
specifying the evidential selection procedure.  The halfer could treat her
current waking experience as being randomly selected from her waking
experiences in the actual world.  In the original two-coins example, by
Bayes' rule this results in $$P(\hbox{HH} | \hbox{see H}) =
\frac{P(\hbox{see H} | \hbox{HH}) P(\hbox{HH})}{P(\hbox{see H} | \hbox{HH})
  P(\hbox{HH})+ P(\hbox{see H} | \hbox{HT}) P(\hbox{HT}) + P(\hbox{see H} |
  \hbox{TH}) P(\hbox{TH})}$$ $$= \frac{1 \cdot (1/4)}{ 1 \cdot (1/4) +
  (1/2) \cdot (1/4) + (1/2) \cdot (1/4)} = 1/2$$ thereby escaping the
Halfer Rule's fatal mistake.  But in the modified (cost-cutting) two-coins
variant, we obtain $$P(\hbox{HH} | \hbox{see H}) = \frac{P(\hbox{see H} |
  \hbox{HH}) P(\hbox{HH})}{P(\hbox{see H} | \hbox{HH}) P(\hbox{HH})+
  P(\hbox{see H} | \hbox{HT}) P(\hbox{HT}) + P(\hbox{see H} | \hbox{TH})
  P(\hbox{TH})}$$ $$= \frac{1 \cdot (1/4)}{ 1 \cdot (1/4) + 1 \cdot (1/4) +
  1 \cdot (1/4)} = 1/3$$ so that this is still a sensible generalization of
halfing.  Still, this generalization is not without its own troubles.  For
one, applying this generalization to the scenario described
by~\cite{Pittard15:When} results in the same credences that he advocates,
which lead to his disagreement paradox.  (Indeed, he argues for these
credences based on a similar evidential selection procedure.)

It seems, then, that generalizing to arbitrary examples will require the
halfer to adopt a rule that leads to one variety or another of unintuitive
consequences.  Perhaps a rule can be found whose unintuitive consequences
are, upon further inspection, quite reasonable, or at least a bullet worth
biting in order to hold on to halfing.  But the so-called Halfer Rule is
not it.  It leads to unacceptable consequences, including egregious
violations of the Reflection Principle -- and this principle is one of the
main motivations for being a halfer in the first place.

\bibliography{beauty}
\bibliographystyle{plainnat}

\end{document}